\documentclass[superscriptaddress,prd,aps,preprint,showpacs]{revtex4}
\newcommand{\bea}{\begin{eqnarray}}
\newcommand{\eea}{\end{eqnarray}}

\begin{document}
\title{On one-loop impacts of the Rashba coupling}

\author{J. R. Nascimento, A. Yu. Petrov}
\affiliation{Departamento de F\'{\i}sica, Universidade Federal da Para\'{\i}ba\\
 Caixa Postal 5008, 58051-970, Jo\~ao Pessoa, Para\'{\i}ba, Brazil}
\email{jroberto, petrov@fisica.ufpb.br}
\author{H. Belich}
\affiliation{Departamento de F\'{\i}sica e Qu\'{\i}mica, Universidade Federal do Esp\'{\i}%
rito Santo, Av. Fernando Ferrari, 514, Goiabeiras, 29060-900, Vit\'{o}ria,
ES, Brazil.}
\email{belichjr@gmail.com}

\begin{abstract}
In this paper, we describe the one-loop contributions in QED with Rashba coupling. We show that all purely nonminimal contributions are explicitly finite, so, the whole theory is one-loop renormalizable.
\end{abstract}

\pacs{11.30.Cp}

\maketitle

\section{Introduction}

The electroweak unification performed through the Higgs mechanism was one of
the largest achievements of the Standard Model of particle physics (SM).
With an idea of spontaneous gauge symmetry violation, the intermediate
bosons corresponding to weak interaction gain mass and the photons continues
to be massless. To complete the success of experimental predictions, in 2013
the detection of Higgs boson was confirmed. However, it is still necessary
to understand why the mass of the Higgs boson is approximately 126 $GeV$,
the problem of the hierarchy of gauge, dark matter, and dark energy. Then we
need a theory that goes beyond the Standard Model.

One proposal of investigation is a line of research that deals with the
Spontaneous Lorentz Violation (SLV) which is induced by fluctuations of
primordial fields in the spacetime which grows up when we aims to arrives at
Planck scale ($10^{19}$ $GeV$). This possibility of SLV allows to introduce
at least one privileged direction in the spacetime \cite{bras, baeta, muon,
meson, meson2, barion2, photon, kost2, electron, neutrino}. The first
attempt to include SLV in the Standard Model of Particle Physics became
known as the Standard Model Extended (SME) \cite{sam, col, coll2}.

In order to verify the possibility of SLV we have the option to observe a
non-trivial background by measurements performed in the particle
accelerators and in low-energy scenarios involving quantum mechanical
effects \cite{bb15, data, manoel, bb3, louzada, manoel2}. Namely the Quantum
Mechanical description of a interactive particle with this background can
give us a hint for this fundamental theory. By a non-minimal coupling with this
environment we can estimate the energy scale at which SLV can emerge through
the bounds calculated from the uncertainty of interference experiments \cite%
{belich, belich1, bb2, bb4, lbb}. This line is based on the idea that these
anisotropies can generate new Berry phases \cite{knut1} acquired by a
particle which is moving in this region \cite{knut2}.

In this paper we follow this line of research describing the
spontaneous violation of the Lorentz symmetry caused by a tensor background.
Particularly, a non-minimal Lorentz-violating coupling which appears in \cite{BaBel, knut3} calls our attention. Using such non-minimal couplings, we
discuss the arising of the Landau system and the influence of a
Rashba-like coupling induced by a Lorentz symmetry violation scenario in the
nonrelativistic quantum dynamics for a spin-$1/2$ neutral particle.
In the ref. \cite{BaBel2}, with an analogue of the Landau system for a
neutral particle, a interesting bound of the Lorentz breaking
term, that is, $gb_{0}<2.2\times 10^{-6}(eV)^{-3}$ is estimated.

In this paper, we intend to generate nonlinear contributions to the action via an appropriate coupling. In the section 2, we carry out the quantum calculations, and the section 3 is our conclusion.

\section{Non-minimal Lorentz-violating coupling}

The Spin-Orbit Coupling (SOC), which is derived from the Dirac equation
by the prescription of the Foldy-Wouthuysen Representation \cite{bb15, data,
manoel, bb3, louzada, manoel2}, has recently been revisited in order to understand a
number of new proposals for materials with unusual behavior. The emergence
of new possibilities of spintronic devices \cite{rash1} without applying a
magnetic field has been the cause nowadays of a intensive research in
Condensed Matter \cite{rash2} and High Energy Physics \cite{knut1,knut2}. The SOC can appear in new materials due to 3 -dimensional
coupling, to the presence of the surface acting as an interface (Rashba and Dresselhaus effect), and due to the 1-dimensional coupling. Our
objective in this article is to call attention to new types of coupling in
electrodynamics \cite{BaBel, knut3, BaBel2} and possible scenarios of their
manifestation. To be more specific, the problem is whether we can generate such couplings by
radiative correction processes. Our objective in this paper is to  study the perturbative impacts of Rashba type couplings.

The possibility to go beyond the SM we use consists, by relaxing the
renormalizability requirement, in proposing an effective Dirac equation with non-minimal
coupling since we are searching for a more fundamental theory \cite{belich1,
aether, aether2}.

Let us consider the spinor QED with the Rashba coupling \cite{BaBel}: 
\begin{equation}
S=\int d^{4}x\bar{\psi}(i\partial \!\!\!/-eA\!\!\!/-\frac{g}{2}F_{\mu \alpha
}F_{\phantom{\alpha}\nu }^{\alpha }\gamma ^{\mu }b\!\!\!/\gamma ^{\nu
}-m)\psi .  \label{rashba}
\end{equation}
Although this theory is non-renormalizable, it displays very interesting properties at the one-loop level. Actually, we see that there is no one-loop divergences generated by the non-minimal coupling in this theory.
We will find the lower contribution to the one-loop effective action. It is
clear that due to the absence of the $\gamma _{5}$ matrix and Levi-Civita
symbol, we cannot have the Carroll-Field-Jackiw term.

The propagator of the spinor is usual, $<\psi(k)\bar{\psi}(-k)>=\frac{1}{k\!\!\!/+m}$. Also, we restrict ourselves by constant background electric
and magnetic fields, thus imposing the condition $\partial_{\mu}F_{\nu\lambda}=0$.

Then, the contribution with only one vertex, of the first order in $b_{\mu}$, 
evidently gives zero result. Indeed, one cannot form a scalar from a product of
tensors with odd total number of indices (one index of $b_{\mu}$ and even number of
indices of any order of $F_{\mu\nu}$). Hence, the lower contribution is 
\begin{eqnarray}
\label{expr}
\Gamma_4=\frac{1}{2}(F^2)_{\alpha\gamma}(F^2)_{\mu\rho}b_{\beta}b_{\nu}{\rm %
tr}\int\frac{d^4k}{(2\pi)^4}\frac{\gamma^{\alpha}\gamma^{\beta}\gamma^{%
\gamma}(k\!\!\!/+m) \gamma^{\mu}\gamma^{\nu}\gamma^{\rho}(k\!\!\!/+m)}{%
(k^2-m^2)^2}.
\end{eqnarray}
Here $(F^2)_{\mu\nu}=F_{\mu\alpha}F^{\alpha}_{\phantom{\alpha}\nu}$. We note the symmetry of $F^2$ tensor, i.e. $(F^2)_{\mu\nu}=(F^2)_{\nu\mu}$. To proceed with the expression (\ref{expr}), we
use the formula 
\begin{eqnarray}
{\rm tr}(\gamma^{\alpha}\gamma^{\beta}\gamma^{\gamma}\gamma^{\delta})=4(%
\eta^{\alpha\beta}\eta^{\gamma\delta}-\eta^{\alpha\gamma}\eta^{\beta\delta}+%
\eta^{\alpha\delta}\eta^{\gamma\beta}),
\end{eqnarray}
and go to $d$-dimensional space-time, making the replacement $k_{\alpha}k_{\beta}\to\frac{k^2}{d}\eta_{\alpha\beta}$.
Also, we keep only even orders in momenta since the odd ones yield zero contributions. We have for the numerator $N^{\alpha\beta\gamma\mu\nu\rho}$: 
\begin{eqnarray}
N^{\alpha\beta\gamma\mu\nu\rho}&=&{\rm tr}[\gamma^{\alpha}\gamma^{\beta}%
\gamma^{\gamma}(k\!\!\!/+m)
\gamma^{\mu}\gamma^{\nu}\gamma^{\rho}(k\!\!\!/+m)=  \nonumber \\
&=& \gamma^{\alpha}\gamma^{\beta}\gamma^{\gamma}k\!\!\!/
\gamma^{\mu}\gamma^{\nu}\gamma^{\rho}k\!\!\!/+m^2\gamma^{\alpha}\gamma^{%
\beta}\gamma^{\gamma} \gamma^{\mu}\gamma^{\nu}\gamma^{\rho}]+\ldots= 
\nonumber \\
&=&{\rm tr}[\frac{k^2}{d}\gamma^{\alpha}\gamma^{\beta}\gamma^{\gamma}%
\gamma_{\sigma}
\gamma^{\mu}\gamma^{\nu}\gamma^{\rho}\gamma^{\sigma}+m^2\gamma^{\alpha}%
\gamma^{\beta}\gamma^{\gamma} \gamma^{\mu}\gamma^{\nu}\gamma^{\rho}]+\ldots,
\end{eqnarray}
where dots are for irrelevant terms. Then, we use the relation 
\begin{eqnarray}
\gamma_{\sigma}\gamma^{\mu}\gamma^{\nu}\gamma^{\rho}\gamma^{\sigma}=-2%
\gamma^{\rho}\gamma^{\nu}\gamma^{\mu},
\end{eqnarray}
which gives 
\begin{eqnarray}
N^{\alpha\beta\gamma\mu\nu\rho}&=&{\rm tr}[-\frac{2k^2}{d}\gamma^{\alpha}%
\gamma^{\beta}\gamma^{\gamma}
\gamma^{\rho}\gamma^{\nu}\gamma^{\mu}+m^2\gamma^{\alpha}\gamma^{\beta}%
\gamma^{\gamma} \gamma^{\mu}\gamma^{\nu}\gamma^{\rho}]+\ldots.
\end{eqnarray}
We remind that this trace should be contracted with $(F^2)_{\alpha%
\gamma}(F^2)_{\mu\rho}b_{\beta}b_{\nu}$, and $(F^2)_{\mu\rho}$ is symmetric.
Actually, it means that the equivalent trace for $N^{\alpha\beta\gamma\mu\nu%
\rho}$, under replacement of $\mu$ by $\rho$ in one of the terms, is 
\begin{eqnarray}
N^{\alpha\beta\gamma\mu\nu\rho}&\simeq&(-\frac{2k^2}{d}+m^2){\rm tr}%
[\gamma^{\alpha}\gamma^{\beta}\gamma^{\gamma}
\gamma^{\mu}\gamma^{\nu}\gamma^{\rho}]+\ldots.
\end{eqnarray}
Then, we consider $\Gamma_4$, which after these replacements takes the form 
\begin{eqnarray}
\Gamma_4=\frac{1}{2}(F^2)_{\alpha\gamma}(F^2)_{\mu\rho}b_{\beta}b_{\nu}{\rm %
tr}\int\frac{d^dk}{(2\pi)^d}\frac{(-\frac{2k^2}{d}+m^2) {\rm tr}%
[\gamma^{\alpha}\gamma^{\beta}\gamma^{\gamma}
\gamma^{\mu}\gamma^{\nu}\gamma^{\rho}]}{(k^2-m^2)^2}.
\end{eqnarray}
Then, it is easy to see that 
\begin{eqnarray}
(F^2)_{\alpha\gamma}b_{\beta}\gamma^{\alpha}\gamma^{\beta}\gamma^{%
\gamma}=2(F^2)_{\alpha\gamma}b^{\alpha}\gamma^{\gamma}-b\!\!\!/
(F^2)_{\alpha}^{\alpha}.
\end{eqnarray}
So, we have 
\begin{eqnarray}
\Gamma_4=\frac{1}{2}\int\frac{d^dk}{(2\pi)^d}\frac{(-\frac{k^2}{2}+m^2) }{%
(k^2-m^2)^2}{\rm tr}\{[2(F^2)_{\alpha\gamma}b^{\alpha}\gamma^{\gamma}-b\!\!%
\!/
(F^2)_{\alpha}^{\alpha}][2(F^2)_{\beta\delta}b^{\beta}\gamma^{\delta}-b\!\!%
\!/ (F^2)_{\beta}^{\beta}]\}.
\end{eqnarray}
Calculating the trace, we find 
\begin{eqnarray}
\Gamma_4&=&\frac{1}{2}\int\frac{d^dk}{(2\pi)^d}\frac{(-\frac{2k^2}{d}+m^2) }{%
(k^2-m^2)^2}[16(F^2)^{\alpha\gamma}b_{\alpha}(F^2)_{\beta\gamma}b^{%
\beta}-4b^2(F^2)_{\beta}^{\beta}(F^2)_{\alpha}^{\alpha}-  \nonumber \\
&-&16(F^2)_{\alpha\beta}b^{\alpha}b^{\beta}(F^2)_{\gamma}^{\gamma}].
\end{eqnarray}
It remains to integrate over the momenta. We do the Wick rotation: 
\begin{eqnarray}
I=\frac{1}{2}\int\frac{d^dk}{(2\pi)^d}\frac{(-\frac{2k^2}{d}+m^2) }{%
(k^2-m^2)^2}=\frac{i}{d}\int\frac{d^dk_E}{(2\pi)^d}\frac{(k^2_E+\frac{d}{2}m^2) }{%
(k^2_E+m^2)^2}
\end{eqnarray}
However, this is nothing more than the miraculous integral from \cite{aether}
which is finite despite naively it involves both quadratic and logarithmic
divergences. For the first manner of calculating this integral, we choose $d=4$. In this  case, similarly to \cite{aether}, we find that 
\begin{eqnarray}
I=-\frac{im^2}{64\pi^2}.
\end{eqnarray}
Within this prescription, we get after the inverse Wick rotation 
\begin{eqnarray}
\Gamma_4=\frac{m^2}{64\pi^2}[16(F^2)^{\alpha\gamma}b_{\alpha}(F^2)_{\beta%
\gamma}b^{\beta}-4b^2(F^2)_{\beta}^{\beta}(F^2)_{\alpha}^{\alpha}
-16(F^2)_{\alpha\beta}b^{\alpha}b^{\beta}(F^2)_{\gamma}^{\gamma}].
\end{eqnarray}
We note that this integral is ambiguous, so, other prescriptions for its
calculation yield other results. 

Within the second manner, we choose $d=4-\epsilon$.
 Afterward, our integral $I$ is 
\begin{eqnarray}
I=\frac{i}{4-\epsilon}\int\frac{d^{4-\epsilon}k_E}{(2\pi)^{4-\epsilon}}\frac{(k^2_E+\frac{4-\epsilon}{2}m^2) }
{(k^2_E+m^2)^2}.
\end{eqnarray}
For a small $\epsilon\neq 0$ this integral is equal to zero, cf. \cite{aether2}. Hence, our
four-point function is ambiguous.

The next correction to be studied involves four vertices (the contributions with three vertices will vanish identically unless we consider derivatives of external
fields). However, it is equivalent to the four-leg diagram of the usual QED
where the replacement of the external legs by the rule $A_{\gamma}\to
2(F^2)_{\alpha\gamma}b^{\alpha}-b_{\gamma} (F^2)_{\alpha}^{\alpha}$ is carried out. And the
four-leg diagram in QED (with no derivatives on external fields) is well
known to vanish. 

We note that the divergent contributions from Feynman diagrams with three (similarly, five, seven etc.) "new" vertices will also vanish -- indeed, it is well known that the one-loop three-point function of the gauge field in QED yields zero result, and also, we can remind that an $n$-point in our theory can be obtained from that one in QED through the replacement $A_{\gamma}\to 2(F^2)_{\alpha\gamma}b^{\alpha}-b_{\gamma}(F^2)_{\alpha}^{\alpha}$, while the integral over momenta is just the same as in QED. In other words, since the one-loop three-point function in QED vanishes in some regularization to provide the gauge symmetry, we can conclude that all divergent contributions from diagrams with three vertices in our theory will vanish as well.

The (one-loop) graphs with six and more "new" vertices are explicitly finite. Actually,
we showed that the theory (\ref{rashba}) is one-loop finite at $e=0$.
We notice that in \cite{BaBel}, namely the case $e=0$ was treated. At the same
time, it is interesting to discuss the one-loop behavior of the complete
theory (\ref{rashba})   involving both $e$ and $g$ which is an extension of
the QED.

First of all, it is clear that the total number of vertices in any one-loop
Feynman diagram with external gauge legs must be even (indeed, each vertex
carries an odd number of indices of fields and $b_{\mu}$ vectors contracted
to some Dirac matrices), and the number of the external $bF^2$ lines should be
even as well, otherwise we should have an odd number of $A_{\alpha}$ legs without derivatives, which is inconsistent with the gauge invariance. Second, since our vertices do not involve derivatives acting
on spinor fields, the upper limit for the degree of divergence of an
arbitrary one-loop graph is $\omega=4-V$, where $V$ is a number of vertices.
Actually, it is even less if some derivatives are transported to external
fields, being $\omega=4-V-2N_d$, where $N_d$ is the number of derivatives
which do not present in vertices from the beginning but arise within the
derivative expansion. It means that the only one-loop superficially divergent graphs
in the theory are those ones with $V=2$ or $V=4$, and, moreover, a
divergent graph with four vertices cannot contain any minimal vertex because, in order
to form the gauge invariant combination, some derivatives must be moved to external fields associated with these vertices which decreases the degree of divergence of the corresponding Feynman diagram (for example, it is well known that the logarithmically divergent contribution from the $A^4$ graph with four minimal vertices is not gauge invariant and hence
vanishes, and, replacing the external legs by the rule $A_{\gamma}\to 2(F^2)_{\alpha\gamma}b^{\alpha}-b_{\gamma}(F^2)_{\alpha}^{\alpha}$, we find that the same situation occurs for $(bF^2)^4$ contribution), and thus the corresponding Feynman diagram becomes finite. Moreover, there is no way to generate the
CFJ term since there is neither $\gamma_5$ matrices nor Levi-Civita symbol in the classical Lagrangian of the theory, so, the quantum corrections should be some contractions of $F_{\mu\nu}$ tensors only. 

Therefore we see that actually, at the one-loop order we can have only
the divergent corrections proportional to $(bF^2)^2$ or $(bF^2)^4$, besides of the
usual $F^2$. At the same time, we note that the $(bF^2)^4$ contribution
vanishes under some regularization prescription, as we already argued, and the $(bF^2)^2$ contribution was showed above to be explicitly finite. So, we see that actually
the only one-loop divergence in this theory is just that one occurring in
the usual QED (see f.e. \cite{Fuji}), and there is no new one-loop
divergences generated by minimal vertices. Actually, we showed that the
theory involving both usual and Rashba-like couplings is one-loop renormalizable in the gauge sector, with the only divergence is that one arising in an usual QED.

\section{Summary}

We considered the Lorentz-breaking extended QED with the additional Rashba coupling. It turns out to be that in the one-loop approximation, this coupling does not generate any new divergences in the gauge sector, other than that one arising in the usual QED, therefore the resulting theory is one-loop renormalizable, and, for $e=0$, even one-loop finite. This allows to treat this coupling as an important ingredient of a possible Lorentz-breaking extended QED, at least if we disregard higher-loop corrections, or treat the gauge field as purely external one, thus restricting ourselves by the fermionic determinant (nevertheless, we note that non-renormalizable field theory models are intensively used even outside of the fermionic determinant context, see f.e. \cite{Brignole}). Moreover, this theory allows to generate the nonpolynomial effective action of the gauge field (that is, the Euler-Heisenberg action), with all terms, at $e=0$, will be explicitly finite.

We note that all studies we carried out here can be applied as well if we
replace $F_{\mu \nu }$ by its dual $\tilde{F}_{\mu \nu }$ in the fermion-vector coupling, as it has been
done in \cite{BaBel2} within the geometrical phase context. So, the theory with
the coupling on the base of $\tilde{F}$ displays the same one-loop
properties as the theory we considered in the paper. In principle, other non-minimal spinor-vector couplings can be also studied in this manner.

{\bf Acknowledgements.}  This work was partially supported by Conselho
Nacional de Desenvolvimento Cient\'{\i}fico e Tecnol\'{o}gico (CNPq). The work by A. Yu. P. has been supported by the
CNPq project No. 303783/2015-0.

\end{document}